\documentclass[aps,prd,twocolumn,groupedaddress]{revtex4}
\usepackage{graphics}% Include figure files
\begin{document}
\newcommand{\be}{\begin{equation}}
\newcommand{\ee}{\end{equation}}

%\preprint{}

\title{Exploring Parameter Constraints on Quintessential Dark Energy: the Pseudo-Nambu Goldstone Boson Model}


\author{Augusta Abrahamse}

\affiliation{}
\author{Andreas Albrecht}
\author{Michael Barnard}
\author{Brandon Bozek}


\date{\today}

\begin{abstract}
We analyze the constraining power of future dark energy experiments
for Pseudo-Nambu Goldstone Boson (PNGB) quintessence. Following the Dark
Energy Task Force methodology, we forecast data for three experimental
``stages'': Stage 2 represents in-progress projects relevant to dark
energy; Stage 3 refers to medium sized experiments; Stage 4 comprises
larger projects. We determine the posterior probability
distribution for the parameters of the PNGB model using Markov Chain
Monte Carlo analysis. Utilizing data generated on a $\Lambda CDM$
cosmology, we find that the relative power of the different data
stages on PNGB quintessence is roughly comparable to the DETF results
for the $w_0-w_a$ parametrization of dark energy. We also generate data based
on a PNGB cosmological model that is consistent with a $\Lambda CDM$
fiducial model at Stage 2. We find that Stage 4 data based on this
PNGB fiducial model will rule out a cosmological constant by at least
$3 \sigma$.
\end{abstract}

%\keywords{}

\maketitle

%\section{}
%\subsection{}
%\subsubsection{}
\section{\label{intro}Introduction}

A growing number of observations indicate that the expansion of the
universe is accelerating. Given our current understanding of physics,
this phenomenon is a mystery. If Einstein's gravity is correct, then
it appears that approximately $70$ percent of the energy density of
the universe is in the form of a ``dark energy''. Although there are
many ideas of what the dark energy could be, as of yet, none of them
stands out as being particularly compelling. Future observations will
be crucial to developing a theoretical understanding of dark energy.

A number of new observational efforts have been proposed to probe the nature of
dark energy, but evaluating the impact of a given proposal is
complicated by our poor theoretical understanding of dark energy.
In light of this issue, a number of model independent methods have
been used to explore these issues (see for example ~\cite{huterer98,
 huterer00}).  Notably, the Dark Energy Task Force (DETF) produced a report
in which they used the $w_o-w_a$ parametrization of the equation of
state evolution in terms of the scale factor, $w(a)= w_o + w_a(1-a)$ ~\cite{DETF06}. The
constraints on the parameters $w_o$ and $w_a$ were interpreted in
terms of a ``figure of merit'' (FoM) designed to quantify the power of
future data and guide the selection
and planning of different observational programs. Improvements in the
DETF FoM between experiments indicate increased sensitivity to
possible dynamical evolution of the dark energy. This is crucial
information, since current data is consistent with a cosmological
constant and any detection of a deviation from a cosmological constant
would have a tremendous impact. There are, however, a number of
questions left unanswered when the dark energy is modeled using
abstract parameters such as $w_0-w_a$ that are perhaps better addressed
with the analysis of actual theoretical models of dark energy.

First of all, the $w_o-w_a$ parametrization is simplistic and not
based on a physically motivated model of dark energy. Although
simplicity is part of the appeal of this parametrization, some of the
most popular dark energy models exhibit behavior that cannot be
described by the $w_o-w_a$ parametrization. The PNGB Quintessence
model considered in this paper, for instance, allows equations of
state that cannot be approximated by the $w_o-w_a$ parametrization.
\begin{figure}[hb]
\centerline{ % 2 figures on top row
\scalebox{0.33}{\includegraphics{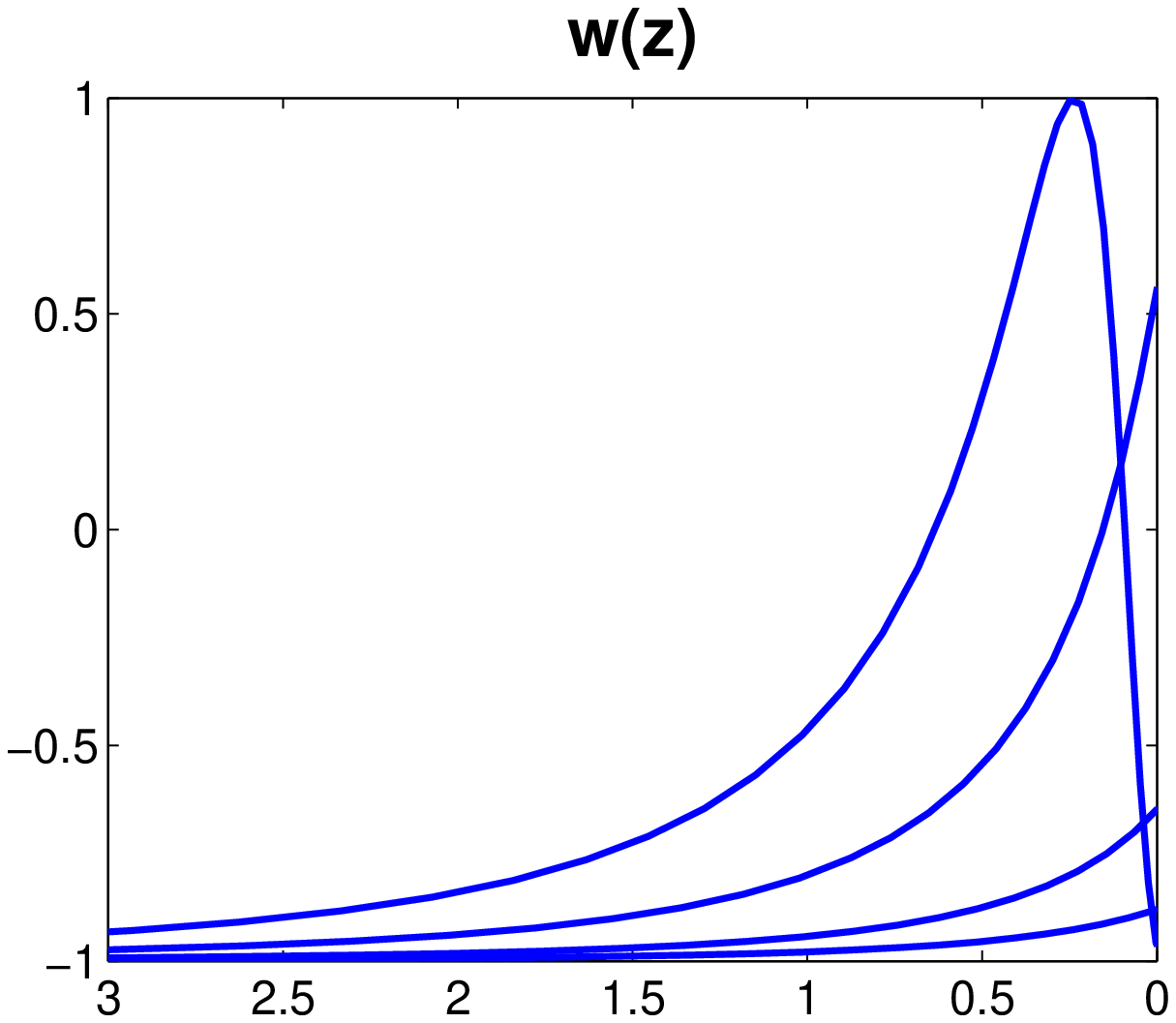}}
\scalebox{0.33}{\includegraphics{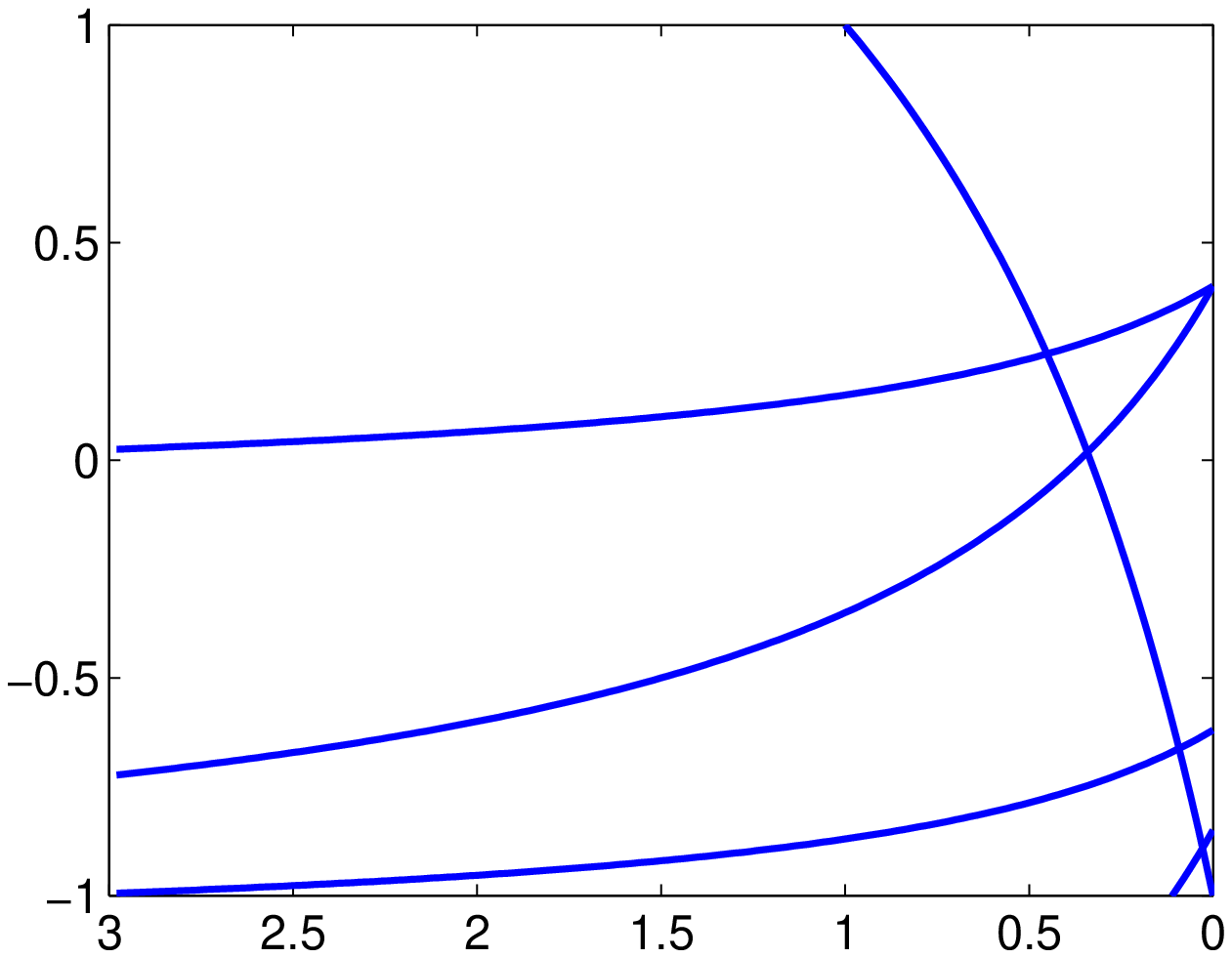}}
}
\caption{\label{wofzs} Examples of possible equations of state
  for PNGB Quintessence (left panel). Attempts to imitate this behavior with $w(z)$
  curves for the $w_0-w_a$ parametrization are depicted on the right.}
\end{figure}

Conversely, the $w_o-w_a$ parametrization may allow solutions that do
not correspond to a physically motivated model of dark energy. Because
of these issues, one could wonder whether the DETF FoM's are somehow
misleading. Various concerns with the DETF analysis have already been
explored. Albrecht and Bernstein used a more complex parametrization
of the scale factor to check the validity of the $w_0-w_a$
approximation~\cite{albrecht06}, and Huterer and Peiris considered a
generalized class of scalar field models
~\cite{huterer06}. Additionally, it has been suggested that the DETF
data models might be improved (see for instance ~\cite{schneider06}).

As of yet, however, no actual proposed models of dynamical dark energy have been
considered in terms of future data. Given the issues above, such an
analysis is an important compliment to existing work. Of course all
specific models of dark energy are suspect for various reasons, and
one can just as well argue that it is better to make the case for new
experiments in a more abstract parameter space rather than tying our
future efforts to specific models that themselves are
problematic. Rather than ``take a side'' in this discussion, our
position is that given the diversity of views on the subject, a
model-based assessment will have an important role in an overall
assessment of an observational program.

In this paper we consider the pseudo-Nambu Goldstone boson
quintessence model of dark energy ~\cite{frieman95}. As one of the
most well-motivated quintessence models from a particle physics
perspective, it is a worthwhile one to study. We use the data models
forecasted by the DETF and generate two types of data sets, one based
on a $\Lambda CDM$ background cosmology and one based
on a background cosmology with PNGB dark energy using a specific
fiducial set of PNGB parameters. We determine the
probability space for the PNGB parameters given the data using Markov
Chain Monte Carlo analysis.  This paper is part of a series of papers
in which a number of quintessence models are analyzed in this
manner\cite{BozekMock,BarnardMock}.

We show that the allowed regions of parameter space shrink as we
progress from Stage 2 to Stage 3 to Stage 4 data in much the same
manner as was seen by the DETF in the $w_0-w_a$ space. This result holds
for both $\Lambda CDM$ and the $PNGB$ data models. Additionally, with
our choice of PNGB fiducial background model, we demonstrate the
ability of Stage 4 data to discriminate between a universe described by a cosmological
constant and one containing an evolving PNGB field.
As cosmological data continues to improve, careful
analysis of specific dark energy models using real data will become
more and more relevant. MCMC analysis can be computationally intensive
and time-consuming. Since future work in this area is likely to
encounter similar challenges, we discuss some of the
difficulties we discovered and solutions we implemented in our MCMC
exploration of PNGB parameter space.

\section{PNGB Quintessence}\label{pngb-model}

Quintessence models of dark energy are popular contenders for
explaining the current acceleration of the
universe~\cite{carroll01,copeland06,sahni02}. Although the
cosmological constant is regarded by many to be the simplest theory of
the dark energy, the required value of the cosmological constant appears to be many
orders of magnitude too small in naive particle theory estimates.
In quintessence models this problem is not solved.  Instead it
is generally sidestepped by assuming some unknown mechanism sets the
vacuum energy to exactly zero, and the dark energy is due to a
scalar field evolving in a pressure dominated state. As such fields can
appear in many proposed ``fundamental theories'', and as the mechanism mimics
ideas familiar from cosmic inflation, quintessence models are regarded
by many (but certainly not by everyone \cite{Bousso:2007gp}) to be at least
as plausible as  a cosmological constant
~\cite{steinhardt05,Albrecht:2007xq}.

Here the quintessence field is presumed to be homogeneous in space,
and is described by some scalar degree of freedom
$\phi$ and a potential $V(\phi)$ which governs the field's
evolution. In an FRW spacetime, the field's evolution is given by
\be \ddot{\phi}+ 3H\dot{\phi}+ \frac{dV}{d\phi} = 0 \label{eq-quintdynam}\ee where
\be H = \frac{\dot{a}}{a}\ee and \be H^2 = \frac{1}{3 M_p^2}(\rho_r + \rho_m + \rho_\phi + \rho_k) \ee
where $M_{P}$ is the reduced Planck mass, $\rho_r$ is the energy density of radiation, $\rho_m$ is the
energy density of non-relativistic matter and $\rho_k$ is the
effective energy density of spacetime curvature.
The energy density and pressure associated with the field are
\begin{eqnarray} \rho_\phi = \frac{1}{2}\dot{\phi}^2 + V(\phi)
  \label{eq-rho}, & P_\phi = \frac{1}{2}\dot{\phi}^2 - V(\phi)
  \label{eq-pressure}\end{eqnarray}
and the equation of state $w$ is given by \be w \equiv \frac{\rho_\phi}{P_\phi}.\ee
If the potential energy dominates the energy of the field, then as can
be seen in Eq. \ref{eq-pressure} the pressure will be negative and in
some cases can be sufficiently so to give
rise to acceleration as the universe expands.

The PNGB model of quintessence is considered compelling because it is one
of the few models that seems natural from the perspective of 4-D
effective field theory. In order to fit current observations, the
quintessence field must behave at late times approximately as a
cosmological constant. It must be rolling on its potential without
too much contribution from kinetic energy, and the value of the
potential must be close to the observed value of the dark energy
density, which is on the order of the critical density of the
universe $\rho_c = 3 H_0^2 m^2_{p} = 1.88 \times 10^{-26} h^2 kg /m^{3}$ or
$2.3 \times 10^{-120}h^2 $ in reduce Planck units, where $h = H_0 / 100$. These
considerations require the field to be nearly massless and the
potential to be extraordinarily
flat from the point of view of particle physics. In general,
radiative corrections generate large mass renormalizations at each
order of perturbation theory unless symmetries exist to suppress this
effect~\cite{kaloper05, kolda98, carroll98}. In order for such fields to seem reasonable, at least on a technical level, their small masses must be protected by symmetries such that when the masses are set to zero they cannot be generated in any
order of perturbation theory. Many believe that Pseudo-Nambu-Goldstone bosons are the simplest way to have
ultra-low mass, spin-0 particles that are natural in a quantum field
theory sense. An additional attraction of the model is that the
parameters of the PNGB model might be related to the fundamental
Planck and Electroweak scales in a way that solves the cosmic coincidence
problem ~\cite{hall05}.

The potential of the PNGB field is well-approximated by
\be V = M^4 [\cos(\frac{\phi}{f}) +1] \label{eq-V}\ee
(where higher derivative terms and instanton corrections are ignored). The evolution of the
dark energy is controlled by the two parameters of the PNGB potential,
$M^4$ and $f$, and the initial conditions, $\phi_I$ and
$\dot{\phi}_I$. We take $\dot{\phi}_I = 0$, since we expect the high expansion
rate of the early universe to rapidly damp nonzero values of
$\dot{\phi}_I$. The initial value of the field, $\phi_I$, takes values
between $0$ and $\pi f$.  This is because the potential is
periodic and symmetric. Since a starting point of $\phi_I/f$ on the
potential is equivalent to starting at $n \pi - \phi_I/f$ and rolling
in the opposite direction down the potential, we require $0< \phi_I/f
< \pi$.

\begin{figure}[hb]
\centerline{ % 2 figures on top row
\label{pngbmodel}
\scalebox{0.33}{\includegraphics{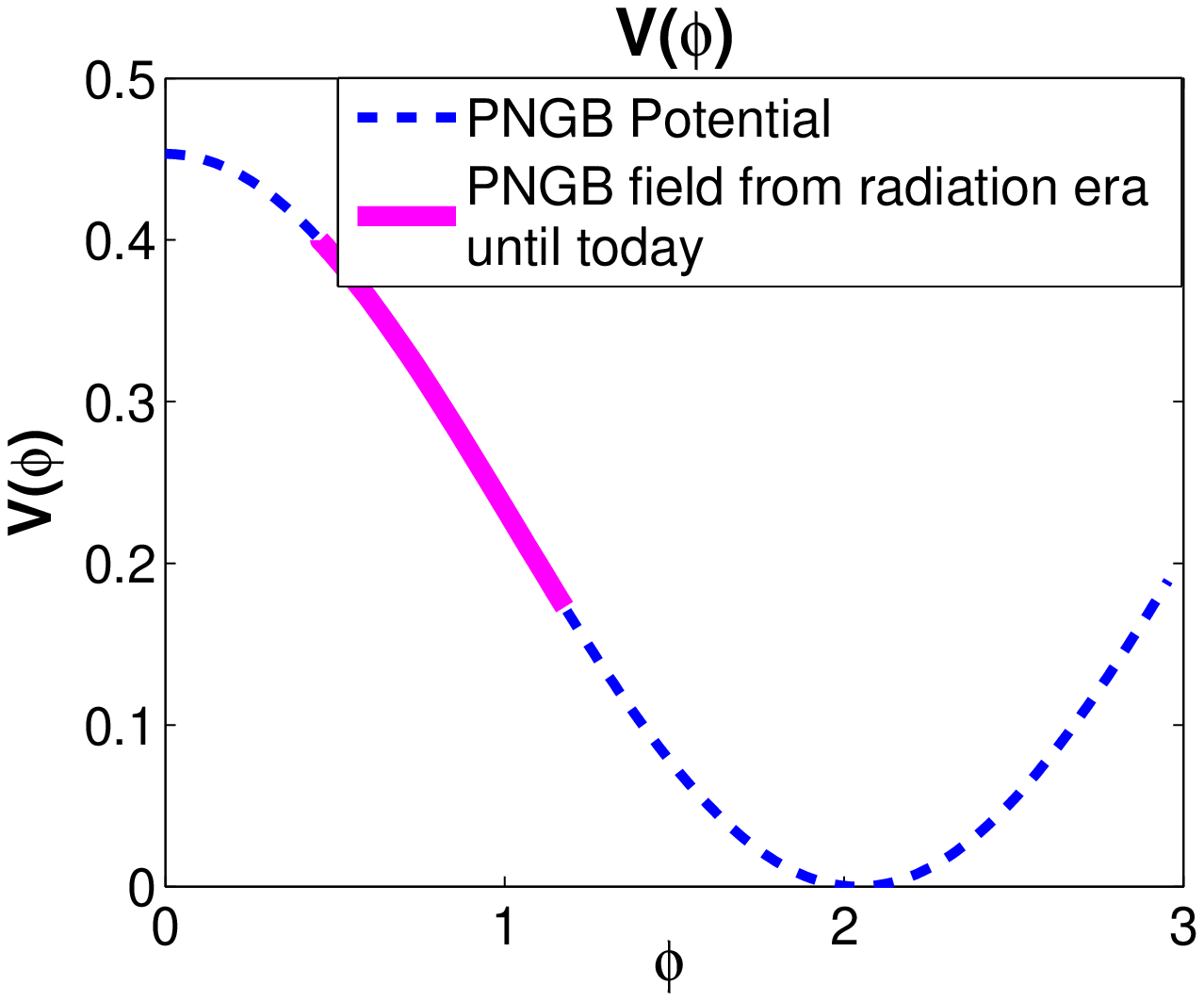}}
\scalebox{0.33}{\includegraphics{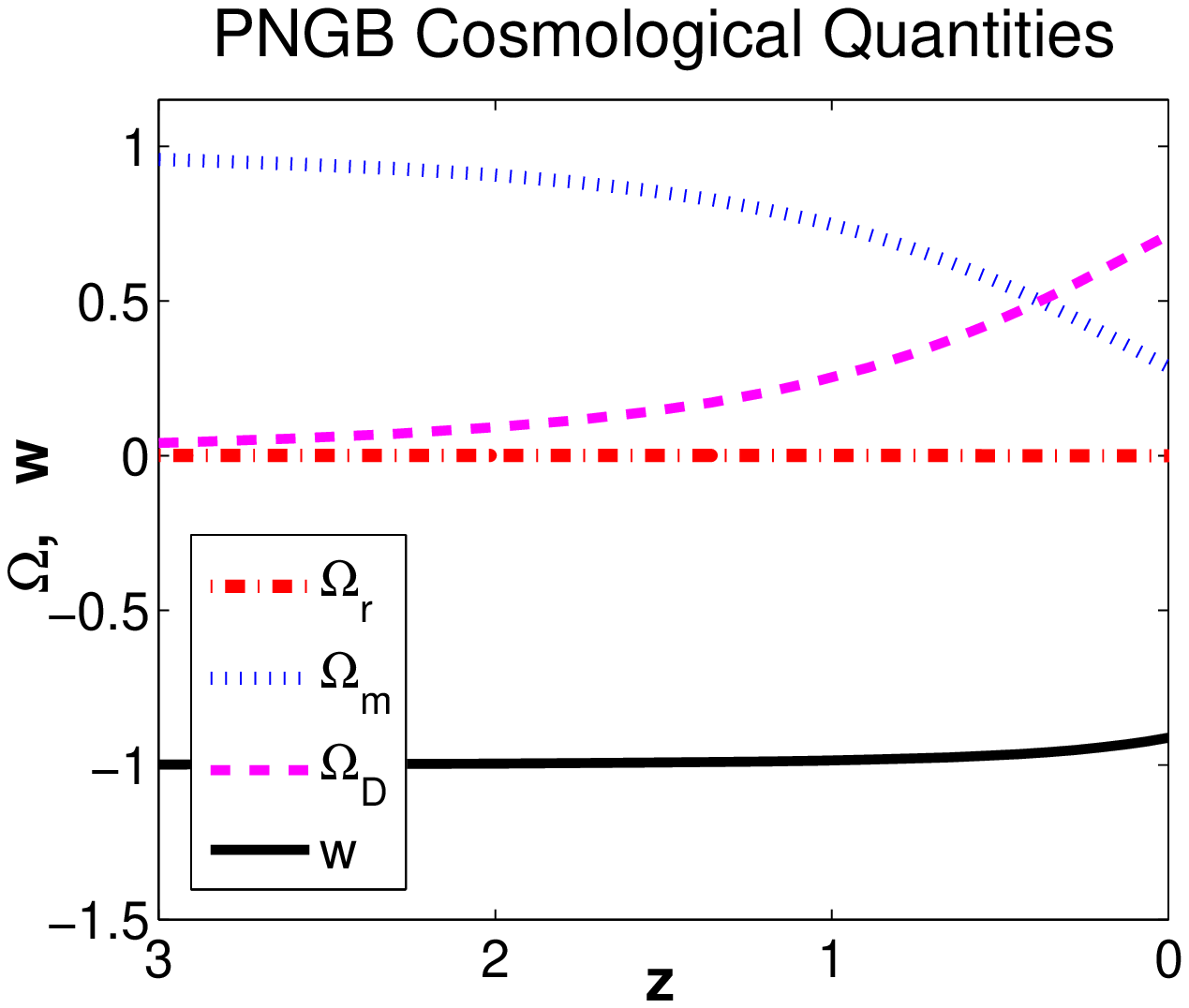}}
}
\caption{\label{pngbexample} An example of a PNGB model and its
resulting cosmological solution for typical values of the PNGB
parameters. The PNGB potential $V(\phi)$ (dashed curve, right panel)
is in units of $h^2$. The evolution of the PNGB field along the
potential since the radiation era is shown by the solid curve
overlaying the potential. The energy densities in
the right panel are related to these units via $\Omega = \omega/h^2$.}
\end{figure}

Additionally, we place the bound $f < M_p$. As will be discussed in the following section, this is
necessary to cut off a divergent direction so that the MCMC chains converge. There are theoretical reasons for this bound as
well. For one, it is valid to neglect higher derivative terms of the PNGB
potential, Eq. (\ref{eq-V}), at least as long as $f < M_p$. In
general, we don't expect to understand 4-D effective field theory at
energies much larger than this. Additionally, there are indications
from string theory that $f$ cannot be larger than $M_p$ ~\cite{dine01, banks03}.

\section{\label{analysis} Analysis and MCMC}

Following the DETF methodology, we generate data sets for future supernova, weak gravitational
lensing, baryon acoustic oscillation and cosmic microwave background
observations. These observations are forecasted for three experimental ``stages'': Stage 2
represents in-progress projects relevant to dark energy; Stage 3
refers to  medium-sized experiments; Stage 4 comprises
larger dark energy projects including  a large
ground-based survey (LST), and/or a space-based program. (Stage I, not considered in our analysis,
represents already completed dark energy experiments, and is less
constraining than Stage 2 data.) ``Optimistic'' and ``pessimistic''
versions of the same simulated data sets give different estimates of
the systematic errors. More information on the specific data models is
given in Appendix \ref{data} (or see also the technical appendix of
the DETF report ~\cite{DETF06}).  In our work we did not use the cluster data because of the
difficulty of adapting the DETF construction to a quintessence
cosmology (the same reasons given in \cite{albrecht06}).

We generate and analyze two versions of data. One is built around a
cosmological constant model of the universe. The other is based on a
PNGB fiducial model. The latter is chosen to be consistent with a
cosmological constant for Stage 2 data.

We use Markov Chain Monte Carlo analysis with a Metropolis-Hastings
stepping algorithm ~\cite{gamerman97,metropolis53, hastings70} to evaluate the
likelihood function for the parameters of our model. The details of
our methods are discussed in Appendix \ref{mcmc}.
MCMC lends itself to our analysis because our probability space is both
non-Gaussian and also depends on a large number of parameters. These
include the PNGB model parameters: $M^4$, $f$, and $\phi_I$, the
cosmological parameters: $\omega_m$, $\omega_k$, $\omega_B$,
$\delta_\zeta$, $n_s$ (as defined by the DETF), and the
various nuisance and/or photo-z parameters accounting for error and
uncertainties in the data).

In order for the results of an MCMC chain to be meaningful, there must
exist a finite, stationary distribution to which the Markov chain may
converge in a finite number of steps. Degeneracies between
parameters, i.e. combinations of different parameters that give rise
to identical cosmologies, correspond to unconstrained directions in
the probability distribution. Unless some transformation of parameters
is found and/or a cut-off placed on these parameters, the MCMC will
step infinitely in this direction and can never converge to a
stationary distribution. Additionally, the shape of the probability
distribution can drastically effect the efficiency of the chain. A
large portion of the task of analyzing the PNGB model, therefore,
involves finding convenient parameterizations and cutoffs to
facilitate MCMC exploration of the posterior distribution.

The probability space of the PNGB model becomes more tractable from an
MCMC standpoint if we transform from the original variables to ones
that are more directly related to cosmological observables constrained
by the data. Such parameterizations make it easier to identify
degeneracies and also tend to make the shape of the
probability distribution more Gaussian. As discussed in
Section \ref{pngb-model}, the dynamics
of the PNGB field depend on its potential $V(\phi) =
M^4(cos(\frac{\phi}{f}) + 1)$, and the specific values of $M^4$, $f$
and $\phi_I$. In order to fit current data the field must hang on the
potential approximating a cosmological constant for most of the
expansion history of the universe. If the field never rolls it acts
as a cosmological constant for all times with a value corresponding
to the initial energy density of the field, $V_I = V(\phi_I)$. To
first order, then, $V_I$ sets the overall scale of the dark energy
density.  Since $V_I$ has more physical significance than $M^4$, it is
a more efficient choice for our MCMC analysis.

\begin{figure}[ht]
\centerline{ % 2 figures on top row
\scalebox{0.33}{\includegraphics{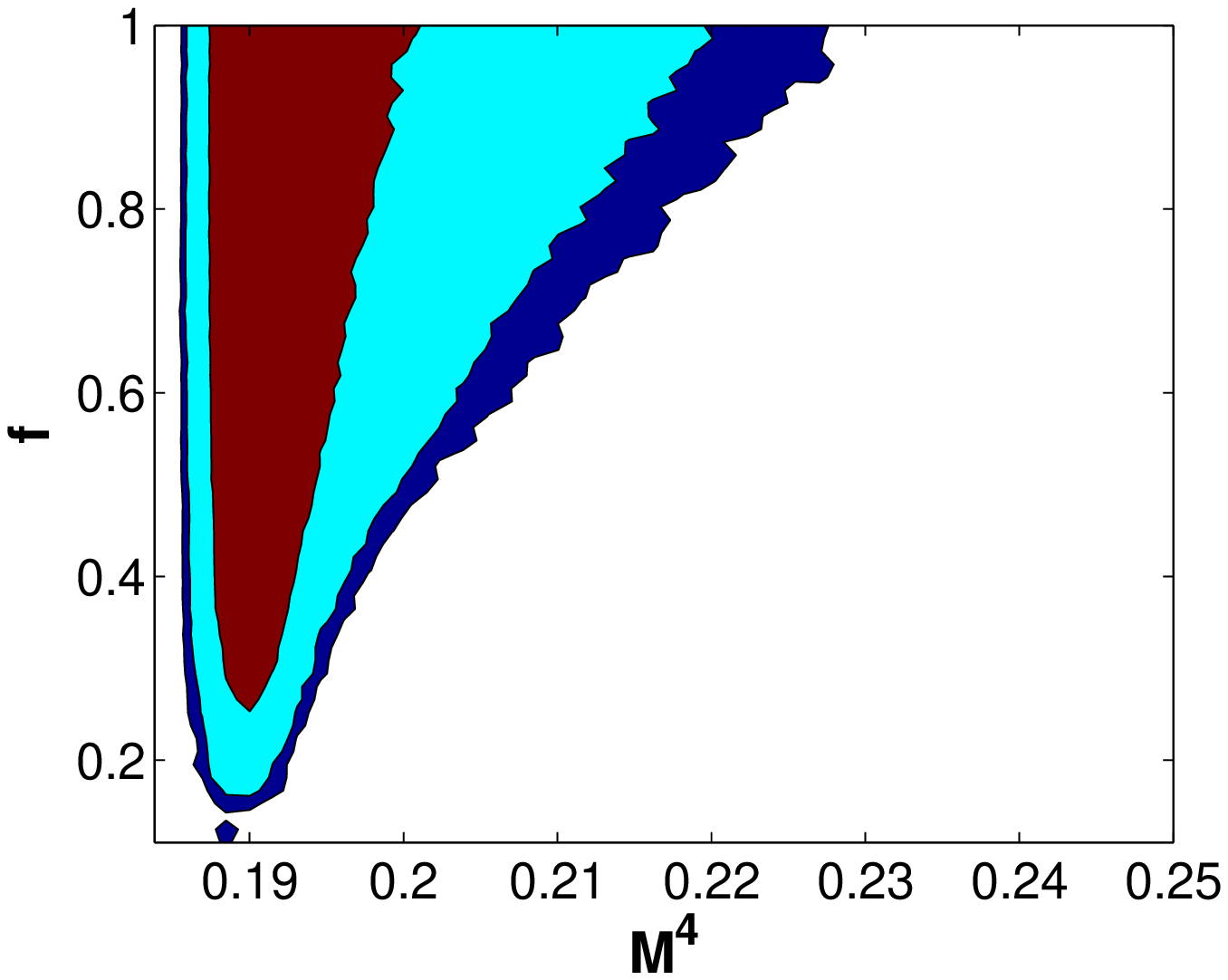}}
\scalebox{0.33}{\includegraphics{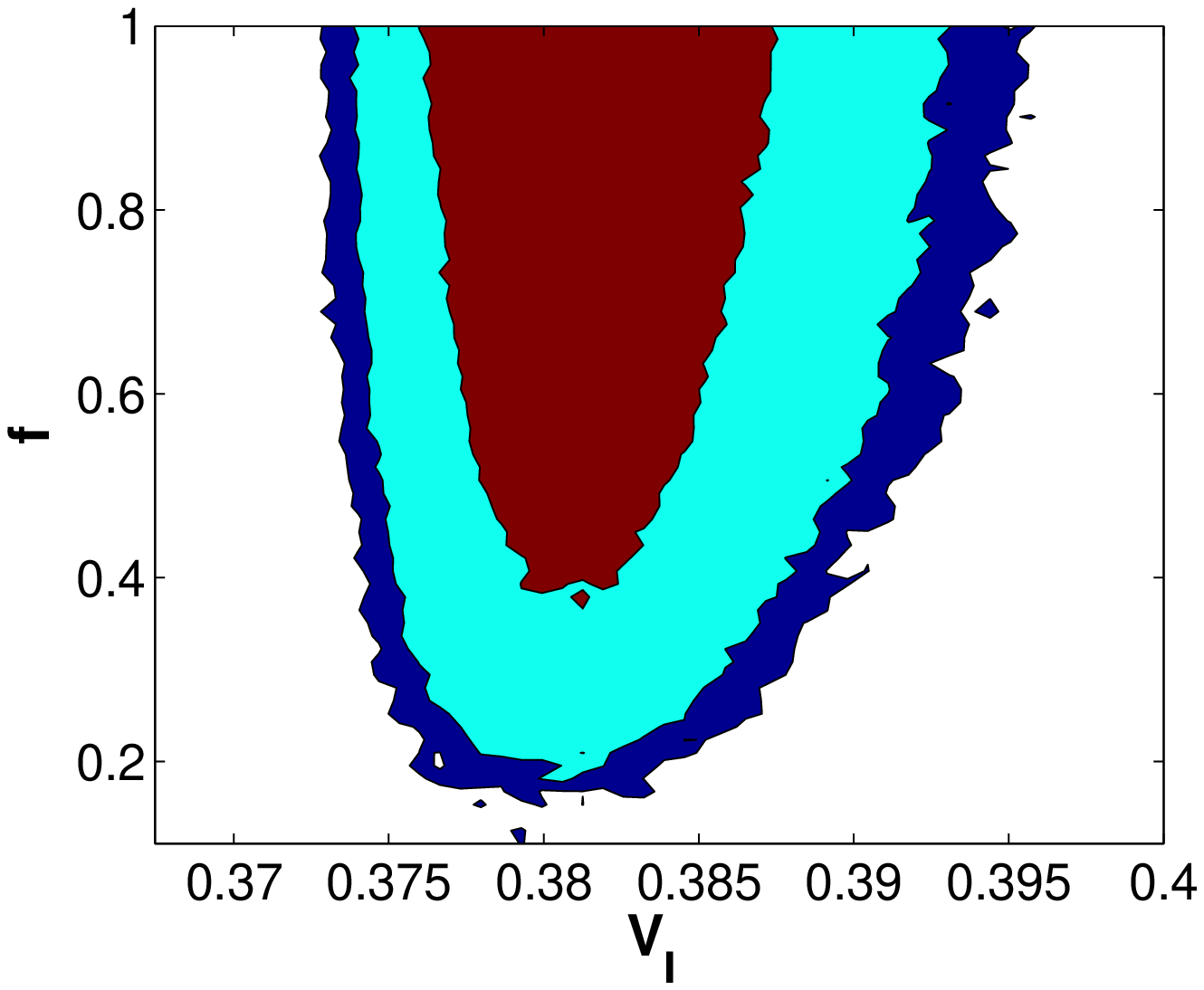}}
}
\caption{\label{vartransform} 2-D confidence regions for $f$ vs. $M^4$
(left panel) and $f$ vs. $V_I$ (right panel). The concave feature of
  the $f$ vs. $M^4$ contours means it is inefficiently explored by
  MCMC.  Contours in the $f-V_I$ space are nearly Gaussian and better
  facilitate convergence.}
\end{figure}

Additionally, the ``phase'', $\phi_I/f$, of the field's starting point in the cosine
potential is closely related to the initial slope of the potential. The slope affects the timescale on which the field
will evolve. If, for instance, $\phi_{I}/f = 0$ the field starts out
on the very top of the potential where the slope is exactly zero, and
the field will not evolve. Starting closer to the inflection point of
the
potential results in a steeper initial slope and the field will roll
faster toward the minimum. Since the variable $\phi_I$ can correspond
to both flat initial slopes (if $f$ is large) or steep ones (if $f$ is
small), $\phi_{I}/f$ is more directly related to the dynamics of
the field and is therefore a superior parameter choice. These new
parameters, as illustrated in Fig. \ref{vartransform}, also result in a
probability distribution that is more Gaussian and thus is more easily
explored by our MCMC algorithm.

Even more important than choosing physically relevant parameters, is
deciding how to handle divergent directions in probability
space. For PNGB quintessence the parameter $f$ must be cut off in some
way because it can become arbitrarily large without being constrained
by the data. For the fiducial model based on a $\Lambda CDM$ universe,
solutions where the field does not evolve for the entire expansion
history of the universe, i.e. behaves as a cosmological constant, can
fit the forecast data perfectly. If such a choice of parameters is found, then
larger and larger values of $f$ will only make the potential flatter
and flatter. If the field did not evolve significantly for the smaller
values of $f$, this will be even more true as the potential
flattens. Hence, $f$ can become arbitrarily large and the
cosmological observables will remain identical.

\begin{figure}[h]
\scalebox{.5}{\includegraphics{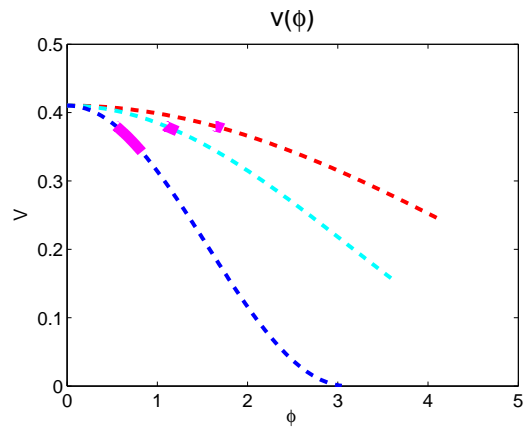}}%
\caption{PNGB potentials (dashed) and with the entire field evolution shown in
  thick solid curves. The different curves show increasing values of $f$ from
  left to right. The smallest value of $f$ (bottom curve) gives a
  nearly static dark energy and fits a cosmological constant well. Any
  larger value for $f$ will also fit the data because the potential
  will be flatter, and the field will evolve even less. \label{fdegeneracy}}
\end{figure}

In general, it is possible to achieve identical deviations from a
cosmological constant by increasing $f$ while at the same time moving
$\phi_I/f$ towards the inflection point of the potential. Increasing
$f$ flattens the potential, but by changing $\phi_I/f$, the slope of
the potential can be held nearly constant and the evolution of the field will not change significantly. In order to achieve results with MCMC, it is necessary to choose some cutoff $f$ so that this infinite
direction is bounded. We choose $f < M_p$ because there is some
theoretical motivation for this choice as detailed in Sec. \ref{pngb-model}.

\section{Results}

\subsection{\label{lambda-fid} $\Lambda CDM$ Fiducial Model}

In this section we present the results of our MCMC analysis for the
combined data sets based on a $\Lambda CDM$ fiducial model. The
model parameters (represented in Table
\ref{lambdafiducialpars}) are in units of $h^2$.

\begin{table}[ht]
\centering
\caption{$\Lambda CDM$ Fiducial Parameter Values (energy densities in units of $h^2$)}
\begin{tabular}{|l l|}
	\hline \hline
$\omega_{DE}$ & $0.3796$ \\
$\omega_{m}$ & $0.146$   \\
$\omega_{k}$ & $0.0$     \\
$\omega_{B}$ & $0.024$   \\
$n_s$        & $1.0 $    \\
$\delta_{\xi}$  & $0.87$ \\
	\hline
\end{tabular}
\label{lambdafiducialpars}
\end{table}

Stage 2 combines supernovae, weak lensing and CMB data. Stages
3 and 4 additionally include BAO data. We marginalize over all but two
parameters to calculate 2-D contours for parameters of the PNGB model
and find the $ 68.27\%$, $ 95.44\%$ and $ 99.73\%$ ($1$, $2$ and $3$ sigma) confidence regions.

\begin{figure}[h]
\scalebox{.5}{\includegraphics{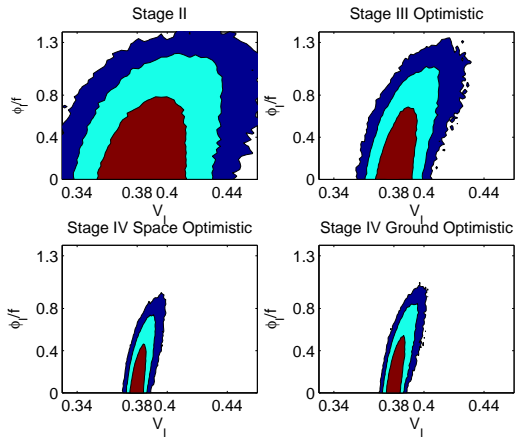}}%
\caption{$V_{I}-\phi_{I}/f$ $1$, $2$ and $3$ sigma confidence regions
for DETF ``optimistic'' combined data
sets. \label{phaseVheight_opt_lambdafid}}
\end{figure}

Fig. \ref{phaseVheight_opt_lambdafid} depicts the contours in the
$V_{I}- \phi_I/f$ plane for Stage 2, and the optimistic versions
of Stage 3, Stage 4 space and Stage 4 LST-ground combined data.  The
horizontal axis where
$\phi_I/f = 0$ corresponds to a cosmological constant. (As explained
above, the field is starting exactly at the top of its potential and
does not roll because the potential is flat at this point.)

\begin{figure}[h t]
\scalebox{.5}{\includegraphics{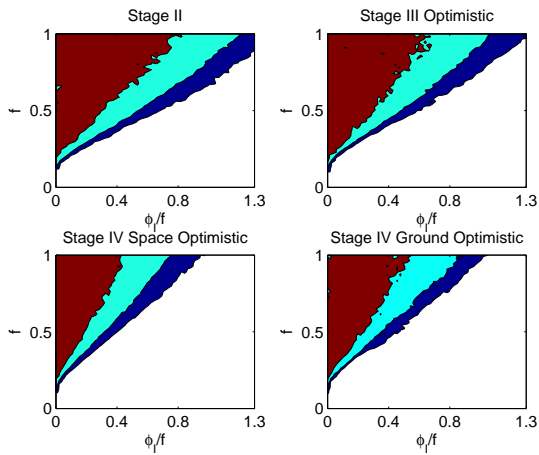}}%
\caption{$f-\phi_{I}/f$ $1$, $2$ and $3$ sigma confidence regions
  for DETF ``optimistic'' combined data sets. \label{fVphase_opt_lambdafid}}
\end{figure}

The value of $V_{I}$ on this axis, therefore, represents the dark
energy density, $\omega_{DE}$, or $\Lambda$.  The contours, as
expected, are centered around $V_{I}=.38$, the fiducial value of
$\omega_{DE}$. It can be seen that the area of the contours shrinks
from Stage 2 to Stage 3 and again from Stage 3 to Stage
4. The shrinking in the $\phi_I/f$ direction roughly corresponds to
constraining deviations from a cosmological constant.
(Although this interpretation is a slight oversimplification, since
for larger values of $f$, $\phi_I/f$ can be non-zero and
perceptible deviations from a cosmological constant will not occur
until sometime in the future.) The reduction in the $V_I$ direction
reflects constraints the data places on the contribution from the
dark energy to the energy density of the universe.

Figs. \ref{fVphase_opt_lambdafid} depicts the $f -\phi_I/f$
contours. Although all values of $f$ are allowed, as $f$
approaches zero the PNGB potential gets narrower, and the phase must start
closer to zero, or else the field will evolve too quickly to
its vacuum state. (The very thin part of the distribution close to
$f=0$ is not resolved by the MCMC analysis.)
For larger values of $f$, $\phi_I/f$ may start further from the peak
of the potential without the field evolving much.
Even for Stage 2, however, the field may not start
past the inflection point of the potential.

Often it is assumed that the PNGB field is initially displaced a small amount from the
potential minimum. But with the constraint we have placed on $f$,
this region of parameter space is no longer accessible for data based
on a $\Lambda CDM$ fiducial model. This is because as the field starts
lower down the slope of the potential, the peak of the potential must
be raised so that $V_I$ may reflect the approximate energy density
needed by the dark energy. But as the peak of the potential gets
higher it also becomes steeper (since $f$ is bounded) and the field
evolves too quickly to fit the data.  It has been suggested, however,
that since we expect quantum fluctuations to displace the field from
the top of potential, it is more reasonable to expect the PNGB field
to start after the inflection point of the potential
~\cite{kaloper05}. If either this argument or the theoretical reasons
for the bound $f < M_p$ could be made more convincing, experimental
results consistent with a cosmological constant could potentially rule
out the PNGB model. As it stands, however, we do not feel the
arguments constraining $f$ and $\phi_I$ are robust enough to make
such a claim.

\begin{figure}[h t]
\scalebox{.5}{\includegraphics{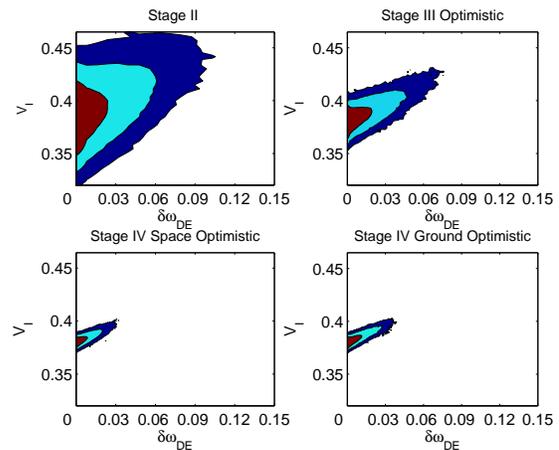}}%
\caption{$V_{I}- \delta \omega_{DE}$ $1$, $2$ and $3$ sigma confidence regions for optimistic
  combined data. Here $\delta \omega_{DE}$ is the amount of change in the dark
  energy density from the radiation era until
  today. \label{heightVdeltaDE_opt_lambdafid}}
\end{figure}

Fig. \ref{heightVdeltaDE_opt_lambdafid} depicts $V_I$ versus
$\delta \omega_{DE}$, where $\delta \omega_{DE} = V_I -
\omega_{DE}(a=1)$. Since PNGB quintessence is a ``thawing'' model of
dark energy\cite{Caldwell:2005tm}, that is, it starts as a cosmological constant until the
field begins to roll causing the amount of dark energy to decrease,
$\delta \omega_{DE}$ reflects the amount the dark energy has deviated
from a cosmological constant. As the DETF found, subsequent stages of
data do better at constraining the evolution of the dark energy. The
fact that Stage 4 space seems a little more constraining than ground
reflects the fact that ground and space data are sensitive to slightly
different features in the dark energy evolution and will be more or
less powerful at different redshifts.  Other quintessence models, such
as the Albrecht-Skordis model ~\cite{BarnardMock} are somewhat better
constrained by the DETF Stage 4 ground data models than by DETF Stage
4 space.

\subsection{PNGB Fiducial Model}\label{pngb-fid}

In addition to considering a $\Lambda CDM$ fiducial model, we evaluate
the power of future experiments assuming the dark energy is really due
to PNGB quintessence.  Our PNGB fiducial parameter values (shown in Table
\ref{pngbfiducialpars} in units of $h^2$) were chosen
such that the fiducial model lies within the $95 \%$ confidence region
for Stage 2 $\Lambda CDM$ data, but demonstrates a small amount of
dark energy evolution that can be resolved by Stage 4 experiments.

\begin{table}[ht]
\centering
\caption{Fiducial Parameter Values (energy densities and $V_I$ in units of $h^2$,$f$ in reduced Planck units}
\begin{tabular}{|l l|}
	\hline \hline
$\omega_{m}$  & $0.145$   \\
$\omega_{k}$  & $0.0$    \\
$\omega_{B}$  & $0.024$  \\
$n_s$         & $1.0$ \\
$\delta_{\xi}$ & $0.87$ \\
$V_{I}$    & $0.4319$ \\
$\frac{\phi_I}{f}$& $0.8726$\\
$f$           & $0.7103$ \\
	\hline
\end{tabular}
\label{pngbfiducialpars}
\end{table}

\begin{figure}[h]
\centerline{ % 2 figures on top row
\scalebox{0.33}{\includegraphics{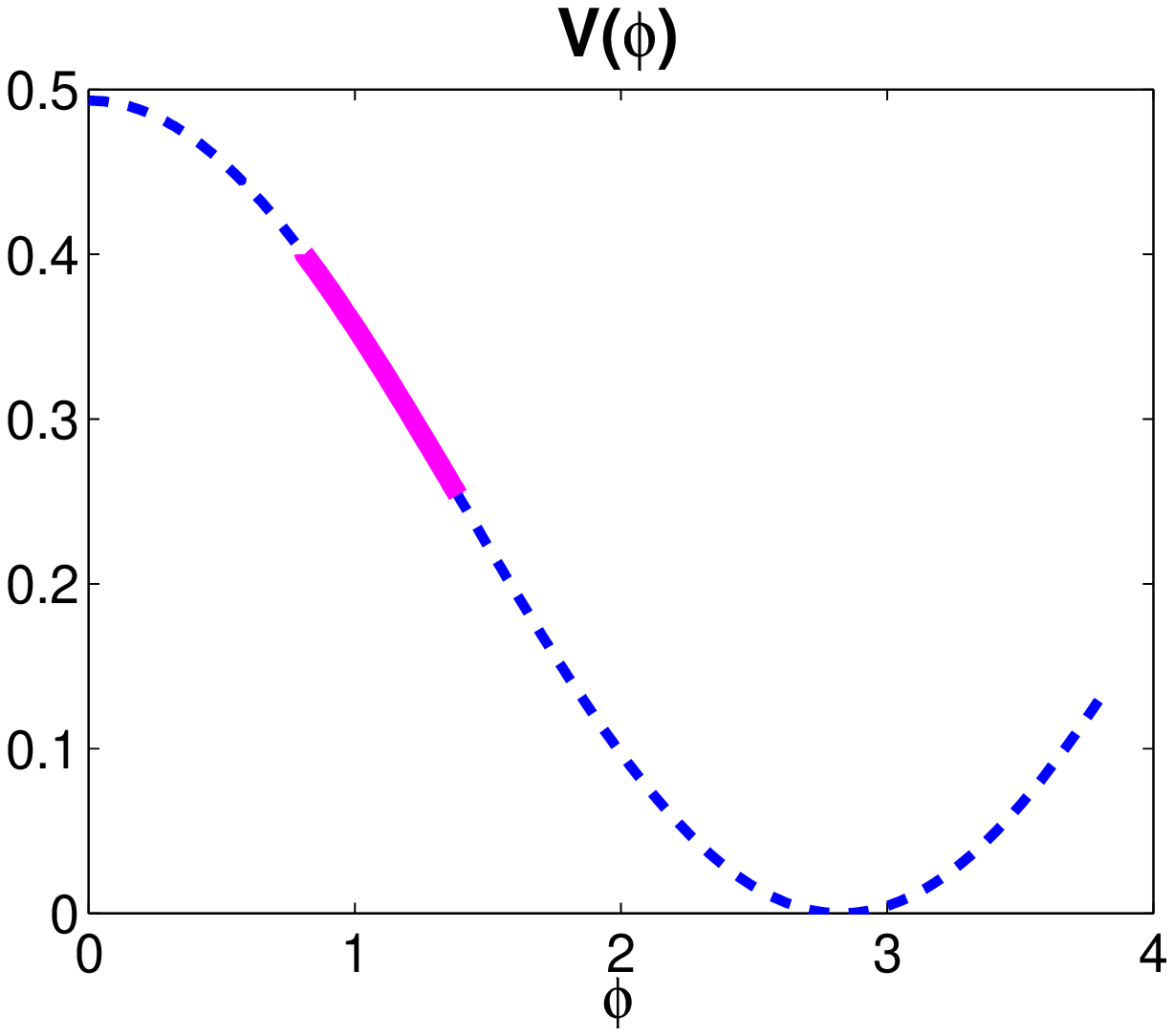}}
\scalebox{0.33}{\includegraphics{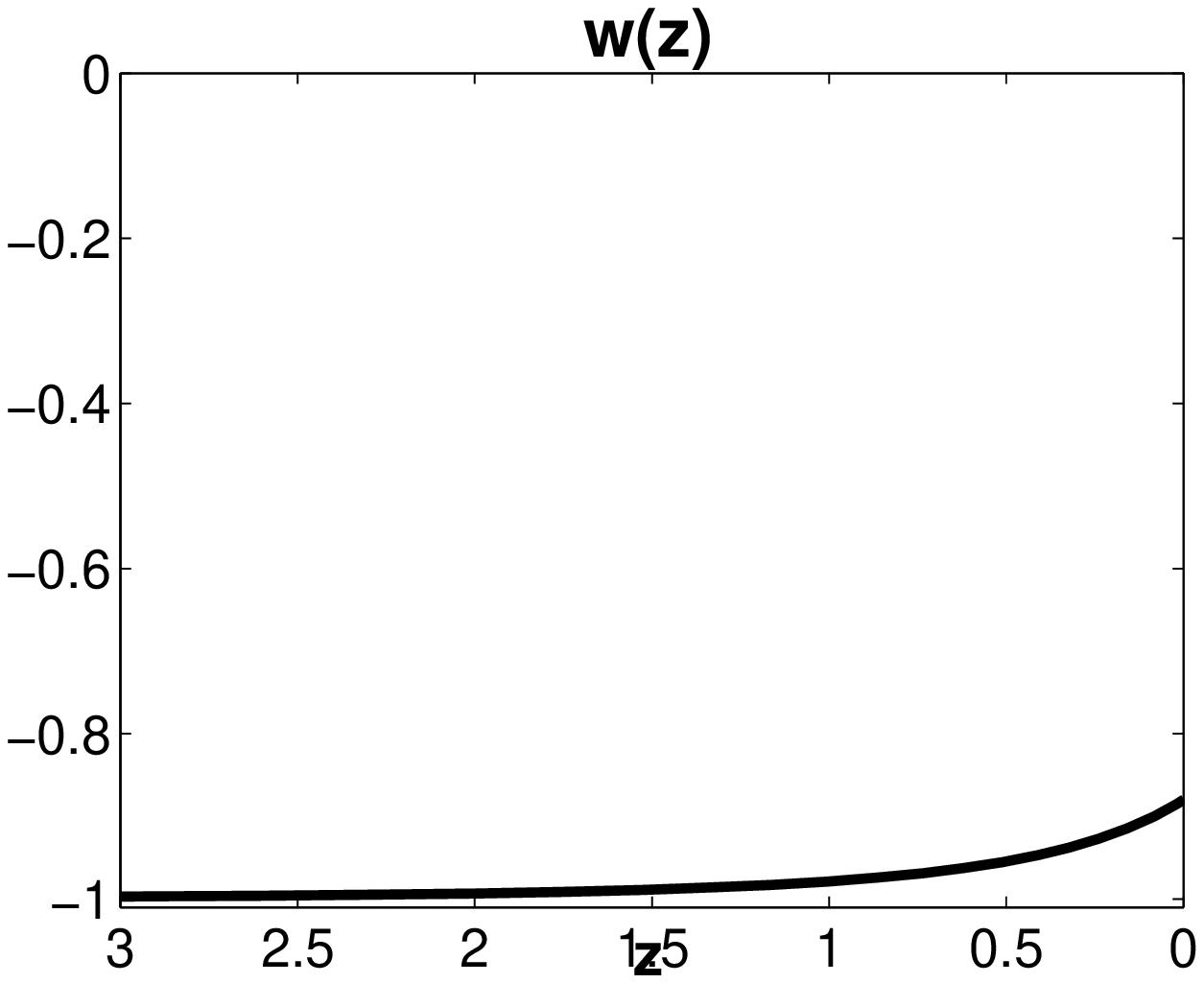}}
}
\caption{\label{pngbpotential} The evolution of
  the PNGB fiducial model field (left panel, solid curve) in the PNGB potential
  (dashed curve).  The corresponding equation of state evolution
  is shown in the right panel.}
\end{figure}

The left panel of Fig. \ref{pngbpotential} shows the potential and evolution of the
field for this model. The right panel depicts $w(z)$. It can be seen
that today ($z=0$) the deviation of the field from $w(z) = -1$ is only about
$ 10\%$.

\begin{figure}[h t]
\scalebox{.5}{\includegraphics{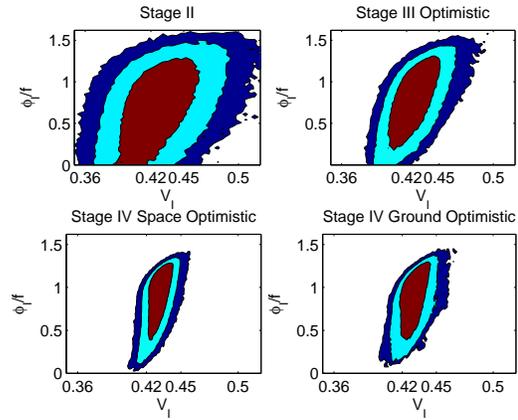}}%
\caption{$V_{I}-\phi_{I}$ $1$, $2$ and $3$ sigma confidence regions
for DETF optimistic combined data sets using the PNGB background
cosmological model.
\label{phaseVheight_opt_pngbfid}}
\end{figure}

\begin{figure}[h t]
\scalebox{.5}{\includegraphics{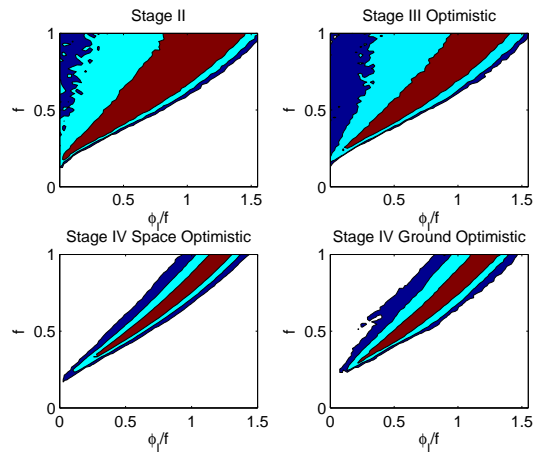}}%
\caption{$f-\phi_{I}/f $ $1$, $2$ and $3$ sigma confidence regions
for DETF optimistic combined data sets using the PNGB background
cosmological model. \label{fVphase_opt_pngbfid}}
\end{figure}

Repeating our MCMC analysis for the PNGB fiducial model, we again
marginalize over all but two parameters to depict the 2-d confidence
regions for the dark energy parameters. Fig.
\ref{phaseVheight_opt_pngbfid} and
depict the $V_I-\phi_I/f$
contours. It can
be seen that the $\phi_I/f = 0$ axis corresponding to the field
sitting on the top of its potential and not evolving, is allowed at Stage 2 but becomes
less favored by subsequent stages of the data. By Stage 4 it
is ruled out by more than $3 \sigma$.

\begin{figure}[h !]
\scalebox{.5}{\includegraphics{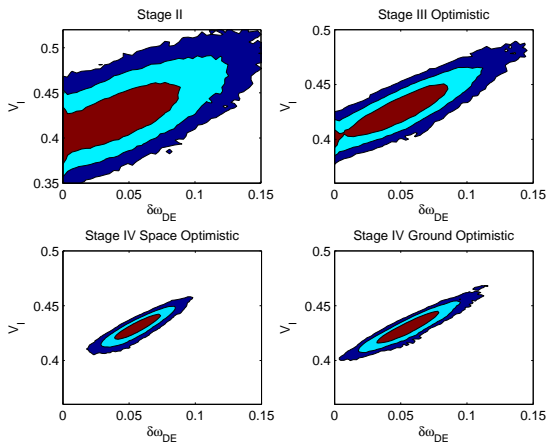}}%
\caption{$V_{I}-\delta \omega_{DE}$ $1$, $2$ and $3$ sigma confidence regions for DETF optimistic combined data,
where $\delta \omega_{DE}$ is the amount of change in the dark
energy density from the radiation era until
today. \label{heightVdeltaDE_opt_pngbfid}}
\end{figure}

Fig. \ref{fVphase_opt_pngbfid} depicts the $\phi_I/f - f$
contours. Again it can be seen that for larger
values of $f$,  $\phi_I/f$ must be non-zero.  By Stage 4 optimistic,
only extreme fine tuning with $f$ allows $\phi_I/f$ to approach zero,
so that the field will be displaced from the top of the potential and
have started to roll by just the right amount by late times.

Fig. \ref{heightVdeltaDE_opt_pngbfid} depict $V_I$ versus $\delta
\omega_{DE}$. At Stage 2 $\delta \omega_{DE}= 0 $ is still is
within the $1 \sigma$ confidence region. But subsequent stages of the
data disfavor this result. Stage 4 optimistic rules out
zero evolution of the dark energy by more than $3 \sigma$.

\section{Discussion and Conclusions}\label{conclusions}

With experiments such as the ones considered by the DETF on the
horizon, data sets will be precise enough to make it both feasible and
important to analyze dynamic models of dark energy.  The analysis of
such models, therefore, should play a role in the planning of these
experiments.

With our analysis of PNGB quintessence, we have shown how future data
can constrain the parameter space of this model. We have shown
likelihood contours for a selection of combined DETF data models, and
found the increase in parameter constraints with increasing data quality
to be broadly consistent with the DETF results in $w_0-w_a$ space.
Direct comparison with the DETF figures
of merit is non-trivial because PNGB quintessence depends on three
parameters, whereas the DETF FoM were calculated on the bases of two,
but in our two dimensional projections we saw changes in the area that
are consistent with DETF results. Specifically, the DETF demonstrated a
factor of roughly three decrease in allowed parameter area when moving
from Stage 2 to good combinations of Stage 3 data, and a factor of
about ten in area reduction when going to from Stage 2 to Stage 4.  We
saw decreases by similar factors in our two dimensional projections.
We have presented likelihood contour plots for specific projected data
sets as an illustration. In the course of this work we produced many
more such contour plots to explore the other possible data
combinations considered by the DETF including the data with
``pessimistic'' estimates of systematic errors.  We found no
significant conflict between our results in the PNGB parameter space
and those of the DETF in $w_0-w_a$ space.

As discussed in \cite{Albrecht:2007xq}, we believe the fact that we have
demonstrated (here and elsewhere  \cite{BozekMock,BarnardMock})
 results that are broadly similar to  those of the DETF despite the
 very different families of functions  $w(a)$ considered is related to
 the fact pointed out in ~\cite{albrecht06} that overall the good DETF data
 sets  will be able to constrain many  more features of $w(a)$ than
 are present in the $w_0-w_a$ ansatz alone.

As data continues to improve, MCMC analysis of dynamic dark energy
models will likely become more popular.  Our experience with the PNGB
model could be relevant to future work.  We find that the
theoretical parameters of the model are not in general the best choice
for MCMC.  Transforming to variables that are closely related to the
physical observables can help MCMC converge more
efficiently.  Additionally, it is necessary to cut off unconstrained
directions in parameter space.  It would be desirable to find bounds
that have some physical motivation.  For PNGB quintessence, we find
that the initial value of the potential, $V(\phi_{I})$, and the
initial ``phase" of the field, $\phi_{I}/f$, are more convenient than
the original model parameters, and that there is some motivation for
placing the bound $f < M_p$.

Finally, we have demonstrated the power Stage 4 data will have for
detecting time evolution of the dark energy.  The PNGB fiducial model
we choose is consistent with Stage 2 data (and with current data by
extension).  If, however, the universe were to in fact be described by
such a dark energy model, then by Stage 4 we would know to better than
$3$ sigma that there is a dynamic component to the dark energy.

\appendix

\section{\label{data} Data}

For each step in the MCMC chain, we integrate numerically to calculate
the theoretical quantities dependent on the dark energy. We start our
integration at early times with $a = 10^{-15}$ and we end the
calculation at $a= 2$.  We compare these value with the observables
generated based on our fiducial models.  With the uncertainties in the
data forecast by the DETF we can calculate the likelihood for each
step in the chain.  What follows is an overview of the likelihood
calculation for each type of observation we consider.

\subsection{Type 1a Supernovae}

After light curve corrections, supernovae observations provide the apparent magnitudes, $m_i$, and the redshift values, $z_i$, for supernova events. The apparent magnitudes are related to the theoretical model through the distance modulus, $\mu(z_i)$, by \be m_i = M + \mu(z_i)\ee where
\be \mu(z_i) = 5 log_{10}(d_l(z_i)) + 25 \ee $M$ is the absolute
magnitude and
\be
d_L (z_i)= \frac{1}{a}\left\{
\begin{array}{l l}
\frac{1}{\sqrt{|k|}}\sinh(\sqrt{|k|}\chi(z_i)) & \quad k<0 \\
\chi(z_i) & \quad k=0 \\
\frac{1}{\sqrt{|k|}}\sin(\sqrt{|k|}\chi(z_i)) & \quad k>0 \\
\end{array} \right.
\ee
and
\be
\chi(z_i) = \eta_0 - \eta(z_i)  \equiv \int^{1}_{a_i} \frac{da}{a^2 H(a)}
\ee
with $|k| = H_0^2|\Omega_k|= (\frac{H_0}{h})^2 |\omega_k|$.

Uncertainties in  absolute magnitude $M$ as well as the absolute scale
of the distance modulus lead to the introduction of an offset $\mu_{off}$
nuisance parameter in all SNe data sets, giving $\mu(z_i) \rightarrow
\mu(z_i) +\mu_{off}$.

Other systematic errors are modeled by more nuisance parameters. The
peak brightness of supernovae, for instance, may have some z-dependent
behavior that is not fully understood. We include this uncertainty in
our analysis by allowing the addition of small linear and quadratic
offsets in z. Additionally, each SNe data model combines a collection
of nearby supernovae with a collection of more distant ones. Possible
differences between the two groups are modeled by considering the
addition of a constant offset
to the near group. The distance modulus becomes
\be \mu(z_i)_{calc} = \mu(z_i) +\mu_{off} + \mu_{lin}z_i +
\mu_{quad}z_i^2 + \mu_{shift}z_{near} \ee.

In addition, some experiments will measure supernovae redshifts
photometrically instead of spectroscopically. There may be a bias in
the measurement of the photo-z's in each bin. This uncertainty is
expressed by another set of nuisance parameters, $\delta z_i$, that
can shift the values of each $z_i$. These observables become $\mu_i =
\mu(z_i + \delta z_i)$.

Priors are assigned to each of the nuisance parameters (except for
$\mu_{off}$, which is left unconstrained) which reflect the projected
strength of the various observational programs. Additionally,
statistical errors are presumed to be Gaussian and are given by the
diagonal covariance matrix $C_{ij} = \sigma^2_{i}$, where
$\sigma_i$ reflects the uncertainty in the $\mu_i$ observables for each
data set.

The likelihood function $L$ for the supernovae data can be calculated from
the chi-squared statistic, where $ \chi^2 \equiv -2 ln L$. For data
sets with photometrically determined redshifts chi-squared is
\begin{eqnarray} \chi^2 &=& \sum(\frac{\mu(z_i) - \mu(z_i)_{data}}{\sigma_i^2}) +
\frac{\mu_{lin}}{\sigma_{lin}^2} + \frac{\mu_{quad}}{\sigma_{quad}^2} \nonumber \\
&+& \frac{\mu_{near}}{\sigma_{near}^2} + \sum(\frac{\delta_i}{\sigma_{z_i}})
.\end{eqnarray}
For data sets with spectroscopic redshifts the chi-squared is the
same minus the contribution from redshift shift parameters.

\subsection{Baryon Acoustic Oscillations}

Large scale variations in the baryon density in the universe has a signature in the matter power spectrum that when calibrated via the CMB provides a standard ruler for probing the expansion history of the universe. The observables for BAO data (after extraction from the mass power spectrum) are the comoving angular diameter
distance, $d_a^{co}(z_i)$, and the expansion rate, $H(z_i)$, where
\be d_a^{co} = ad_L \ee and $z_i$ indicates the z bin for each data
point. The quality of the data probe is modeled by the covariance
matrix for each observable type, as described in section 4 of the
DETF technical appendix. Additionally, some BAO observations use
photometrically determined redshifts, in which case $\delta z_i$ are
added as nuisance parameters as for the supernovae, to
describe the uncertainty in each redshift bin.

The likelihood function for BAO observations is
\begin{eqnarray} \chi^2 & = &  \sum ( \frac{d_a^{co} (z_i) - d_{a-data}^{co}
(z_i)}{\sigma_i^2} + \frac{H(z_i) - H_{data}(z_i)}{\sigma_i^2}) \nonumber \\
&   + &\sum \frac{\delta z_i}{\sigma_{z_i}} \end{eqnarray}

\subsection{Weak Gravitational Lensing}

Light from background sources is deflected from a straight path to the
observer by mass in the foreground. From high resolution imaging of large numbers of galaxies, it is possible to detect statistical correlations in the stretching of galaxies, ``cosmic shear". From this foreground mass distributions can be determined. The mass distribution as a
function of redshift provides a probe of the growth history of density
perturbations, $g(z)$, where  $g(z)$ (in linear perturbation theory)
depends on dark energy via
\be \ddot{g} + 2H\dot{g} =
\frac{3 \Omega_m H_o^2}{2a^3} g .
\ee
Additionally, because the amount
of lensing depends on the ratios of distances between the observer,
the lens and the source, gravitational lensing also probes the
expansion history of the universe, $D(z)$.

The direct observables of weak lensing surveys considered by the DETF
are the power spectrum of the lensing signal and the cross-correlation
of the lensing signal with foreground structure. Systematic and
statistical uncertainties are described by a Fisher matrix in this
space. As is detailed in the DETF appendix, it is possible to
transform from this parameter space to the variables directly
dependent on dark energy, $g(z)$ and $D(z)$. These become the weak
lensing observables we use in our analysis.

In addition to depending on the dark energy model, weak lensing
observations depend on the cosmological parameters $\omega_m$, matter
density, $\omega_B$, Baryon density, $\omega_k$, effective curvature
density, $n_S$, the spectral index, and $lnA_s$, the amplitude of the
primordial power spectrum. These parameters are treated as nuisance
parameters with priors imposed by the Fisher matrix.

Lastly, since ground-based lensing surveys will photometrically
determine redshifts, as for SNe and BAO data, we must model the
uncertainty in redshift bins. Again this is done by allowing each
$z_i$ bin to vary by some amount $\delta z_i$.

The weak lensing observables are given be the vector
\be \overrightarrow{X}_{obs}=[(\omega_m, \omega_k, \omega_B, n_s,
\delta_\zeta), d_a^{co}(z_i), g(z_i), ln(a(z_i)] \ee
where $a(z_i)$
is the scale factor corresponding to the redshift bins for the data.
The error matrix is non-diagonal in the space of these observables so
chi-squared is given by \be \chi^2 = (\overrightarrow{X}_{obs}-
\overrightarrow{X}_{obs-data}) \mathbf{F_{lensing}}
(\overrightarrow{X}_{obs}- \overrightarrow{X}_{obs-data})^\top \ee

\subsection{Planck CMB}

As with baryon oscillations, observations of anisotropies in the
cosmic microwave background probe the expansion history of the universe
by providing a characteristic length scale at the time of last
scattering. As with weak lensing, our Planck observables are
extrapolated from the CMB temperature and polarization maps. The
observable space constrained by Planck becomes: $n_s$,
$\omega_m$, $\omega_B$, $\delta_{\zeta}$, $ln(\Theta_S)$. These
variables are constrained via the Fisher matrix in this space (we use
the same one used in \cite{albrecht06}). The
chi-squared is calculated as
\be \chi^2 = (\overrightarrow{X}_{obs}- \overrightarrow{X}_{obs-data})
\mathbf{F_{Planck}} (\overrightarrow{X}_{obs}-
\overrightarrow{X}_{obs-data})^\top \ee

\section{\label{mcmc} MCMC}
Markov Chain Monte Carlo simulates the likelihood surface for a set of parameters by
sampling from the posterior distribution via a series of
random draws. The chain steps semi-stochastically in parameter space
via the Metropolis-Hastings algorithm such that more probable values
of the space are stepped to more often. When the chain has converged
it is considered a ``fair sample" of the posterior distribution, and the density
of points represents the true likelihood surface. (Explanations of this
technique can be found in ~\cite{gamerman97, bridle02,
christensen01}).

With the Metropolis-Hastings algorithm, the chain starts at an arbitrary position $\theta$ in parameter space. A candidate position $\theta'$ for the next step in the chain is drawn from a proposal probability density $q(\theta,
\theta')$. The candidate point in parameter space is accepted and
becomes the next step in the chain with the probability
\be
\alpha(\theta, \theta') = min \{ 1 , \frac{ P(\theta') q(\theta', \theta) }{P(\theta)q(\theta, \theta')}\}
\ee
where $P(\theta)$ is the likelihood of the parameters given the
data. If the proposal step $\theta'$ is rejected, the point $\theta$
becomes the next step in the chain. Although many distributions are
viable for the proposal density $q(\theta, \theta')$, for simplicity
we have chosen to use a Gaussian normal distribution. (It should be noted that, in general, the dark energy parameters of the model are not Gaussian distributed. The power of the MCMC procedure lies in the fact that it can probe posterior distributions that are quite different from the proposal density $q(\theta, \theta')$.) Since this is symmetric, $q(\theta, \theta')
= q(\theta', \theta)$, we need only consider the ratios of the
posteriors in the above stepping criterion.

For the results of the Markov chain to be valid, it must equilibrate,
i.e. converge to the stationary distribution. If such a distribution
exists, the Metropolis-Hastings algorithm guarantees that the chain will
converge as the chain length goes to infinity. In practice, however,
we must work with chains of finite length. Moreover, from the standpoint of
computational efficiency, the shorter our chains can be and still
reflect the true posterior distribution of the parameters, the
better. Hence a key concern is assuring that our chains have
equilibrated. Though there are many convergence diagnostics, chains
may only fail such tests in the case of non-equilibrium; none
guarantee that the chain has converged ~\cite{carlin96}. We therefore
monitor the chains in a variety of ways to convince ourselves that
they actually reflect the underlying probability space.

Our first check involves updating our proposal distribution $q(\theta,
\theta')$, which we have already chosen to be Gaussian normal. Each
proposal step is drawn randomly from this distribution. The size of
the changes generated in any given parameter direction depend on the
covariance matrix we use to define $q(\theta, \theta')$. We start by
guessing the form of the covariance matrix and run a short chain
($O(10^5)$ steps) after which we calculate the covariance matrix
of the Markov chain. We then use this covariance matrix to define the
Gaussian proposal distribution for the next chain. We repeat this
process until the covariance matrix stops changing systematically.
This implies that the Gaussian approximation to the posterior has been
found. In addition to indicating convergence, this also assists the
efficiency of our chains. The more the proposal distribution reflects
the posterior, the quicker the Markov chain will approximate the
underlying distribution.

One convergence concern is that we might not be exploring the entire probability
space. It is possible, for instance, that if we started our
chains at a random point in probability space and our step sizes are
too small, the chain may have wandered to a local maximum from which
it will not exit in a finite time. We could be missing other features of the
underlying probability space. We convince ourselves that this is not
the case by starting chains at different points in parameter space.
We find that the chains consistently reflect the same probability
distribution, and hence we conclude that we are truly sampling from
the full posterior.

After we have determined our chains are fully exploring probability
space and we have optimized our Gaussian proposal distribution, we
run a longer chain to better represent the probability space of our
variables. We consider the chain to be long enough when the 95\%
contour is reasonably smooth. For most data sets, chains of $O(10^6)$
are sufficient although the larger the probability space, the longer the chains must be. (In
particular, Stage 4 ground data involves a large number of nuisance
parameters and may take two to three times longer to return smooth
contours.) With the final chains, we must control for both burn-in
and correlations between parameter steps. Burn-in refers to the
number of steps a chain must take before it starts sampling from the
stationary distribution. Because we have already run a number of
preliminary chains, we know approximately the mean parameters of our
model. We find that the means refer to a point in probability space
close to the maximum of the distribution. (Generically, this
does not have to be true if the probability space is asymmetrical.)
If we use this as our starting point, our chains do not have to
wander long before they appear to sample from the stationary
distribution. We control for this by removing different amounts from
the start of the chain.  For instance, if we cut out the first $1000$
steps and calculate the contours and compare this to contours
calculated with the first $100000$ steps removed we find that the
shape of the $2-D$ contours remain essentially the same. We can
conclude, therefore, that chains very quickly begin sampling the
posterior distribution and we need not worry about burn-in.

Correlations between steps may also effect the representativeness of
the samples generated via MCMC. The effects, however, may be
controlled for by either thinning the chains by a given amount or by
running chains of sufficient length such that the correlations become
unimportant. We experiment with different thin factors (taking every
step, every $10th$ step and every $50th$ step and we find very little
difference in our results. Hence we conclude that
the sampling of our chains are not greatly effected by correlations.

Lastly, we apply a numerical diagnostic similar to that used by Dick
et al. ~\cite{dick06} to test the conversion of our chains. (This
technique is a modification of the Geweke diagnostic
~\cite{Geweke92}.) We compare the means
calculated from the first $10\%$ of the chain (after burn-in of 1000)
to the means calculated from the last $10\%$. If the chain has
converged to the stationary distribution, then these values should be
approximately equal. If $ \frac{mean_1(\theta_i) -
mean_2(\theta_i)}{\sigma_{ii}}$ is large, where $\sigma_{ii}$ is the
standard deviation determined by the chain for the parameter
$\theta_i$, then the chain is likely to still be drifting. We find
that for our chains $ \frac{mean_1(\theta_i) -
  mean_2(\theta_i)}{\sigma_{ii}}< .1 $ for $95\%$ of the
parameters. The remaining parameters are no less than
$\frac{\sigma_{ii}}{5}$ away from each other. Coupling this with the
qualitative monitoring of
the chains described above, we are confident that our chains do a good
job of reflecting the posterior probability distribution of our
model.

\begin{acknowledgements} We thank Matt Auger, Lloyd Knox, and Michael Schneider for useful discussions and constructive criticism. Thanks also to Jason Dick who provided much useful advice on MCMC. We thank the Tony Tyson group  for use of their computer cluster, and in particular Perry Gee and Hu Zhan for expert advice and computing support. Gary Bernstein provided us with Fischer matrices suitable for adapting the DETF weak lensing data models to our methods, and David Ring and Mark Yashar provided additional technical assistance. This work was supported by DOE grant DE-FG03-91ER40674 and NSF grant AST-0632901.

\end{acknowledgements}

\bibliography{pngbBib}

\end{document}